\documentclass[aps,pra,twocolumn,showpacs,amsmath,amssymb,preprintnumbers,superscriptaddress,10pt]{revtex4-1}

\usepackage{graphicx}
\usepackage{SIunits}
\usepackage{hyphenat}
\usepackage[bookmarks=false]{hyperref}
\usepackage{color}
\usepackage[capitalise]{cleveref}
\crefname{section}{Sec.}{Secs.}
\Crefname{section}{Section}{Sections}
\usepackage[normalem]{ulem} 

\usepackage{ulem}

\begin{document}
	\title{Enhancing secure key rates of satellite QKD  using a quantum dot single-photon source }
	
	\author{Poompong~Chaiwongkhot}
	\email{poompong.ch@gmail.com}
	\affiliation{Institute for Quantum Computing, University of Waterloo, Waterloo, ON, N2L~3G1 Canada}
	\affiliation{Department of Physics and Astronomy, University of Waterloo, Waterloo, ON, N2L~3G1 Canada}
	\affiliation{Department of Physics, Faculty of Science, Mahidol University, Bangkok 10400, Thailand}
	
	\author{Sara~Hosseini}
	\email{sara.hosseini@uwaterloo.ca}
	\affiliation{Institute for Quantum Computing, University of Waterloo, Waterloo, ON, N2L~3G1 Canada}
	\affiliation{Department of Physics and Astronomy, University of Waterloo, Waterloo, ON, N2L~3G1 Canada}
	
	\author{Arash~Ahmadi}
	\affiliation{Institute for Quantum Computing, University of Waterloo, Waterloo, ON, N2L~3G1 Canada}
	\affiliation{Department of Physics and Astronomy, University of Waterloo, Waterloo, ON, N2L~3G1 Canada}
	\affiliation{Walter Schottky Institute, Technische Universität München, 85748 Garching, Germany}
	
	\author{Brendon~L.~Higgins}
	\affiliation{Institute for Quantum Computing, University of Waterloo, Waterloo, ON, N2L~3G1 Canada}
	\affiliation{Department of Physics and Astronomy, University of Waterloo, Waterloo, ON, N2L~3G1 Canada}
	
	\author{Dan~Dalacu}
	\affiliation{National Research Council of Canada, Ottawa, K1A 0R6, Ontario, Canada}
	
	\author{Philip~J.~Poole}
	\affiliation{National Research Council of Canada, Ottawa, K1A 0R6, Ontario, Canada}
	
	\author{Robin~L.~Williams}
	\affiliation{National Research Council of Canada, Ottawa, K1A 0R6, Ontario, Canada}
	
	\author{Michael~E.~Reimer}
	\affiliation{Institute for Quantum Computing, University of Waterloo, Waterloo, ON, N2L~3G1 Canada}
	\affiliation{Department of Electrical \& Computer Engineering, University of Waterloo, Waterloo, ON N2L~3G1  Canada}
	
	\author{Thomas~Jennewein}
	\email{thomas.jennewein@uwaterloo.ca}
	\affiliation{Institute for Quantum Computing, University of Waterloo, Waterloo, ON, N2L~3G1 Canada}
	\affiliation{Department of Physics and Astronomy, University of Waterloo, Waterloo, ON, N2L~3G1 Canada}
	
	\date{\today}
	
	\begin{abstract}
		Global quantum secure communication can be achieved using quantum key distribution (QKD) with orbiting satellites. Established techniques use attenuated lasers as weak coherent pulse (WCP) sources, with so-called decoy-state protocols, to generate the required single-photon-level pulses. While such approaches are elegant, they come at the expense of attainable final key due to inherent multi-photon emission, thereby constraining secure key generation over the high-loss, noisy channels expected for satellite transmissions. In this work we improve on this limitation by using true single-photon pulses generated from a semiconductor quantum dot (QD) embedded in a nanowire, possessing low multi-photon emission ($<10^{-6}$) and an extraction system efficiency of $-15~\text{dB}$ (or 3.1\%). Despite the limited efficiency, the key generated by the QD source  is greater than that generated by a WCP source under identical repetition rate and link conditions representative of a satellite pass. We  predict that with realistic improvements of the QD extraction efficiency to $-4.0~\text{dB}$ (or 40\%), the quantum-dot QKD protocol  outperforms WCP-decoy-state QKD by almost an order of magnitude. Consequently, a QD source could allow generation of a secure key in conditions where a WCP source would simply fail, such as in the case of high channel losses. Our demonstration is the first specific use case that shows a clear benefit for QD-based single-photon sources in secure quantum communication, and has the potential to enhance the viability and efficiency of satellite-based QKD networks.
	\end{abstract}
	
	\maketitle
	
	\section{Introduction}
	
	Quantum key distribution (QKD)~\cite{gisin2002,scarani2009} such as the seminal BB84 protocol~\cite{bennett1984} generates unconditionally secure keys between two distant parties by transmitting encoded photons. While terrestrial implementations have limited range due to inherent channel losses, implementations with orbiting satellites can extend the range of QKD across the globe~\cite{Liao:2017aa,PhysRevLett.120.030501}. However, key generation is diminished by high transmission loss, noise, limited contact time, and imperfections~\cite{bourgoin2013,polnik2020}. In particular, when employing a weak coherent pulse (WCP) source, the secure key length is hampered to account for multi-photon emissions which leak information~\cite{hwang2003, ma2005, bourgoin2015}, to the extent that a given pass of a QKD satellite might not yield any usable secure key. Because the cost of a QKD platform in orbit is substantial and only offers limited access time to a ground station, improvements to the rate of secure key generation are crucial for the platform's viability. While channel performance is limited by telescope apertures and atmospheric quality, the use of true single-photon sources to eliminate information leakage could improve satellite QKD transfer. This is now feasible given recent advances of single-photon emitter devices. Here, we study QKD  using single-photon pulses generated by a semiconductor quantum dot (QD)~\cite{A.WhiteQD} 
	embedded in a photonic nanowire (see, e.g., \cite{dalacu2019}) that ensures efficient and directional light extraction\cite{reimer2016} and as well as high pulse rate and purity \cite{Bottomup2012,Nano2012}.
	Such devices could be excellent candidates for a ground-to-satellite uplink implementation, such as  the Canadian QEYSSat mission~\cite{jennewein2014}, where the sizeable cryogenics and pump lasers required to operate the QD are located on ground. 
	
	While the development of single-photon sources using semiconductor quantum dots has progressed steadily in recent years, their benefit for QKD has yet to be clearly demonstrated. Previous proof-of-concept studies demonstrated that QKD with QD sources could generate keys up to the sifting step~\cite{Intallura2007, heindel2012, Takemoto2015}. Here, we theoretically and experimentally compare the QKD performance of a QD source and a WCP source, focusing on a regime of high channel loss and including finite-size effects, where statistical estimates possess significant uncertainty (and thereby impact secure key length) due to small sample sizes. We estimate secure key rates using the formalism of Ref.~\onlinecite{cai2009} for QD QKD, after taking practical coupling losses with the QD into account~\cite{renner2005,koenig2007}, and the decoy-state model of Refs.~\onlinecite{hwang2003,ma2005,bourgoin2015,curty2010} for WCP QKD.  Note that, to provide a fair comparison of the two sources, we model our calculations assuming the same pulse repetition rates, as well as the same realistic conditions for satellite-based QKD links, including channel losses of 25 to $35~\text{dB}$, fly-by pass duration of 100~s, and background photon noise rate of several 100~Hz.   We believe our study is the first of its kind that demonstrates  a quantum dot single-photon source can substantially enhance the performance of a BB84 QKD protocol under such satellite-link conditions.


	\section{Experiment setup}
	
	\begin{figure*}
		\includegraphics{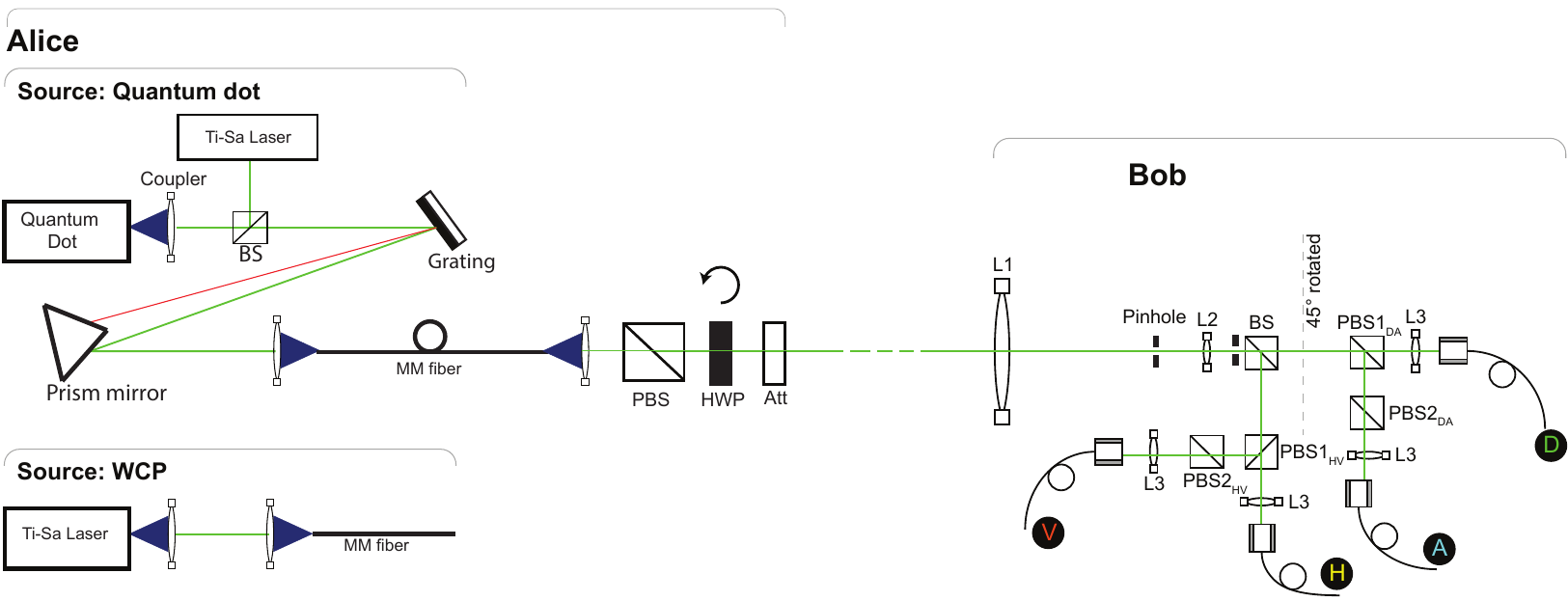}
		\caption{Experimental apparatus. For QD QKD, the QD is excited with a mode-locked titanium:sapphire (Ti:sapph) laser. A grating and a wedge mirror are used to separate exciton pulses from bi-exciton pulses. The photons are then sent to the QKD system. For WCP QKD, photon pulses are sent directly from the Ti:sapph laser to the QKD system. At Alice, the signal polarization is first cleaned up with a polarizing beam splitter (PBS) and then encoded through a motorized half-wave plate (HWP). Attenuators (Att) are used to simulate channel loss, as well as to select intensity levels for the decoy-state WCP protocol. The signal is sent through a free-space quantum channel and measured by a passive basis choice polarization-encoding QKD receiver at Bob (see text for details).}
		\label{fig:setup}
	\end{figure*}
	
	Our experimental apparatus is shown in \cref{fig:setup}. The optical source, QD or WCP, is coupled into multi-mode fiber. The transmitting party, Alice, utilizes a polarizer and a half-wave plate in a motorized rotating stage to encode one of four equally-distributed linear polarizations---horizontal (H), diagonal (D), vertical (V), or anti-diagonal (A)---onto each photon pulse from the selected source. The photons then pass through attenuators emulating channel loss before passing through the free-space channel to the receiver. The  receiving party, Bob, uses a beam splitter and polarizing beam splitters to discriminate the four polarization states with passive choice of measurement basis---the passive-basis-choice design is generally favourable for a satellite payload as it avoids active elements which are more likely to fail in orbit~\cite{bourgoin2015}.
	
	Alice's photons  enter Bob's receiver through a focusing lens (L1, diameter $50~\milli\meter$, focal length $250~\milli\meter$), and a collimating lens (L2, diameter $5~\milli\meter$, focal length $11~\milli\meter$). The collimated beam of $\lesssim 2~\milli\meter$ diameter then passes through a 50:50 beam splitter (BS), and polarization beam splitters ($\text{PBS1}_\text{HV(DA)}$ and $\text{PBS2}_\text{HV(DA)}$) are placed in HV(DA) arm (PBS2 suppresses noise due to comparatively low polarization extinction ratio in the reflected path of PBS1). Lenses (L3) in each of the four polarized outputs focus the beams into four multi-mode fibers connected to single-photon detectors. The times of detection pulses are recorded using a time-tagging system.
	
	Link parameters are chosen following the study of satellite QKD in Ref.~\onlinecite{bourgoin2013}. For each source, we record a series of pulses in each channel transmission loss value, one polarization at a time, for 100~s each, which is a typical usable link time of a satellite in low Earth orbit. The key generation rate is determined using the average observed statistics of each polarization.

	\section{Quantum Dot QKD}
	
\begin{figure}
		\includegraphics[height=7cm]{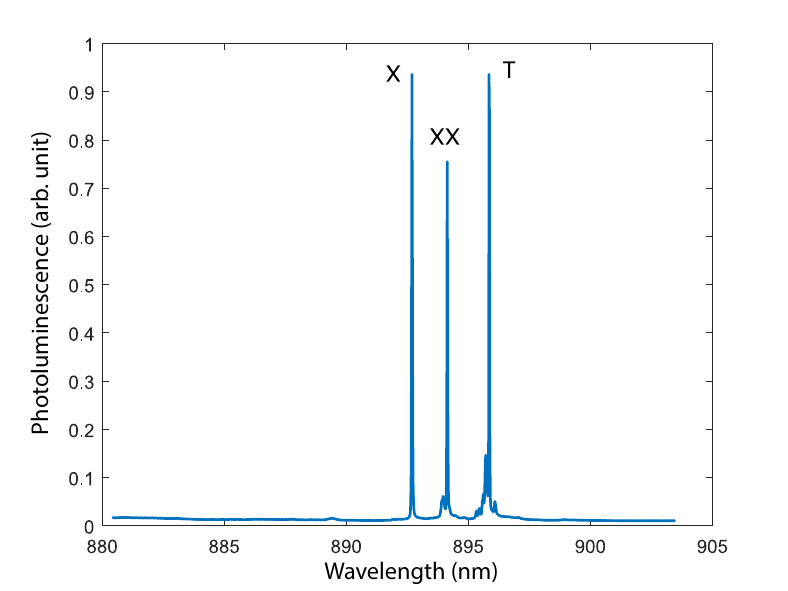}
		\caption{Observed emission spectrum of the QD excited at 830~nm, i.e., above bandgap non-resonant excitation. The spectrum shows three peaks attributed to the exciton $(X)$, biexciton $(XX)$, and charged exciton or trion ($T$). The spectrum is measured by an imaging spectrometer using a 1200~grooves/mm grating.}
		\label{fig:QD_emission}
	\end{figure}

	We utilize a single-photon signal from a wurtzite indium arsenide phosphide (InAsP) quantum dot, embedded in a tapered Indium Phosphide (InP) nanowire \cite{Bottomup2012,Nano2012}. Non-resonant, or incoherent, pulsed pumping is used to excite the quantum dot, effectively releasing carriers above the bandgap of the wurtzite-InP nanowire bandgap transition.
	The photoluminescence spectrum of the quantum dot, \cref{fig:QD_emission}, is captured under off-resonance excitation by 830~nm laser pulses from a titanium:sapphire (Ti:sapph) mode-locked laser at 420~nW power and 76.4~MHz repetition rate. The quantum dot emits exciton photons at 892.67~nm and biexciton photons at 894.2~nm. To separate these two emission lines, we send the quantum dot emission to a polarization-independent transmission grating with 1504 grooves per millimetre. The photons from the excitonic emission are coupled to a multi-mode optical fiber and sent to the QKD state preparation apparatus---these are chosen as they have a higher rate than those from the biexciton emission (see \cref{fig:QD_emission}).  The QD source has an internal loss of about $15~\text{dB}$ due to imperfect photon generation and collection, resulting in an effective pulse rate of 2.6~MHz. 
	
	For QKD security analysis we assume that the phases of each photon pulse are independent, which is a good approximation for QD single photon pulses \cite{jayakumar2013}, while any residual phase could be erased by randomizing the pump phase \cite{Lee:2018aa}. 
	A secure implementation of a QKD source requires fast, random polarization encoding. This could be achieved by multiple methods, including direct \cite{Lee:2018aa} or on-chip phase modulation, fast external electro-optical phase modulators (typical insertion loss of $4~\text{dB}$ at repetition rates of 10~GHz \cite{OEspace-phase}) or  passively combining four QDs with dedicated polarization orientations and switching of the excitation laser pulse to address a random QD at each time slot (insertion loss of $3~\text{dB}$, repetition rates at GHz levels). For our demonstration we use slow variation of the polarization encoding based on a motorized wave-plate.

	
	
\begin{figure}[h]
		\includegraphics{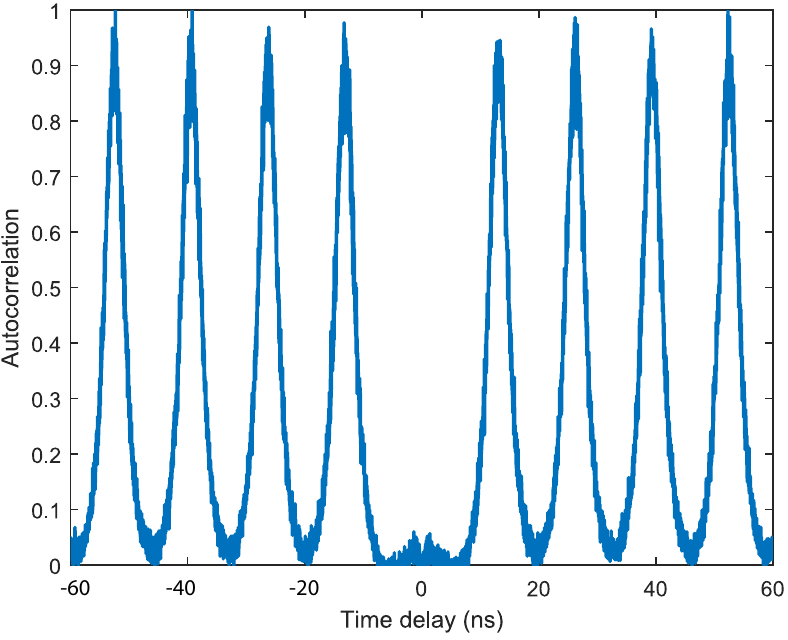}
		\caption{Measured autocorrelation histogram of photon emission from the QD. The data are presented without any background corrections.}
		\label{fig:G2_NRE_TPE_2}
	\end{figure}

	A low  multi-photon emission probability is most critical for a secure QKD implementation. Impressively, the nanowire quantum dot source has a measured second-order correlation  $g^2(0) \approx 0.015$ when excited off-resonance (see \cref{fig:G2_NRE_TPE_2}). Although in semiconductor quantum optics there is a special emphasis placed on the measurement of the $g^2(0)$, recently it has been shown that $g^2(0) < 1/2$  does not provide the exact probability of single- or multi-photon emission \cite{PG2019} but only suggests a non-zero single-photon contribution in the quantum state of the light. This means that even at a low  $g^2(0)$ the source may emit a small fraction of multi-photon pulses, which could permit an adversary to perform a photon number splitting attack and potentially gain information about the key.
	
	To suppress information leakage due to multi-photon emission from the QD, we use the key rate equation for BB84 QKD with an imperfect photon source~\cite{lutkenhaus2000, gottesman2004},
	\begin{equation} \label{eq:finite}
	\begin{aligned}
	L_\text{QD} \quad \geq & \quad nq A\left(1-H\left(\frac{\tilde{E}}{A}\right)-fH(E)-\Delta\right),
	\end{aligned}
	\end{equation}
	where $n$ is the number of raw key bits,  $q = 1/2$ is the sifting ratio, and $E$ is the observed quantum bit error ratio (QBER). $\tilde{E}=E+\frac{1}{2}\sqrt{\{2\ln(1/\varepsilon_\text{PE})+2\ln(m+1)\}(1/m)}$ takes into account the chance that the QBER estimated from a sifted key of size $m=qn$ deviates from the actual value~\cite{renner2005a,cai2009,scarani2008a}, and $\varepsilon_\text{PE}$ is the probability that such deviation occurs. $H(x)=-x\log_2(x)-(1-x)\log_2(1-x)$ is the binary Shannon entropy, and $fH(E)$ is information leakage during error correction with error correction code efficiency $f$. A correction term,
	\begin{equation}
	\label{eq:decoy-delta}
	\Delta = -7m\,\sqrt[]{\frac{1}{m}\log_2\frac{2}{\tilde{\varepsilon}}}-2\log_2\frac{1}{\varepsilon_\text{PA}}-\log_2\frac{2}{\varepsilon_\text{EC}}
	\end{equation}
	accounts for statistical deviations due to finite-size effects~\cite{curty2010,hwang2003,ma2005,bourgoin2015}, with security parameter $\varepsilon = \tilde{\varepsilon}+\varepsilon_\text{PA}+\varepsilon_\text{EC}$. In this experiment, we choose $\varepsilon_\text{EC} = 10^{-10}$, and $\tilde{\varepsilon}$ and $\varepsilon_\text{PA}$ are numerically optimized for the key size under the constraint $1- \varepsilon_\text{EC}>\tilde{\varepsilon} > \varepsilon_\text{PA} \geq 0$. The correction term $A = (p_\text{det}-P_m)/p_\text{det}$ accounts for an adversary's information due to multi-photon pulses \cite{lutkenhaus2000, gottesman2004}, where $p_\text{det}$ is the probability of detection and $P_m$ is the probability of a multi-photon pulse generated by Alice per time slot. Because the photon number distribution of the quantum dot emission is not precisely known, and certainly cannot be presumed to follow that of the coherent state, we employ an alternative method to establish an upper bound for $P_m$.


	
	First,  from the QKD data recorded in Bob's apparatus, we  determine  the likelihood  for a three-way coincidence detection, where more than two detectors `click' within the same time window (5~\nano\second), to be less than $10^{-9}$. This probability is similar to accidental coincidences caused from background noise in the channel and dark counts from the detectors, and implies that source contributions of three of more photons is negligible---we thus do not consider those further.
	
	The  remaining two-photon contributions are characterized  with the help of  a 50:50 beam splitter, where each output is  coupled to an APD, at coupling efficiency $\eta_t = 10\%$ and APD detection efficiency of $\eta_d = 60\%$.  The experiment is run for a duration of 10 hours to obtain sufficient probability of coincident clicks, $C$, in a 5~ns window. With emission of $i$-photon Fock-states at probabilities given by $p_i$, the probability for a 2-fold coincidence is
	\begin{align}
		C &= \frac{1}{2} p_2 \eta^2 +\mathcal{O}(\eta D)+\mathcal{O}(D^2) \\
		&= N_C / N,
		\label{eq:C}
	\end{align}
	where the detection efficiency of the testing device is given by $\eta = \eta_t\eta_d$, $D$ is the dark count probability, $N_C$ is the total number of coincident detections, and $N$ is the number of time slots during 10 hours of data collection. We similarly also determine the probability of `solitary' events $S$ where only one detector clicks within the window,
	\begin{align}
		S &= p_1 \eta + p_2 \eta (\frac{3}{2} - \eta)+ \mathcal{O}(D)\\
		&= N_S/N,
		\label{eq:S}
	\end{align}%
	where $N_S$ is number of solitary detections over the data collection period. In this setup the probability of dark count per detection event is lower than $10^{-7}$ per detection. Thus, the contribution of dark counts to $C$ and $S$ is negligible. By combining \cref{eq:C} and \cref{eq:S}, we find
	\begin{equation}
	p_2 = \frac{2 \kappa p_1}{\eta-3\kappa+2\kappa\eta} ,
	\end{equation} 
	where $\kappa \equiv C / S$ is calculated from the measurement data. Under the assumption that higher photon terms can be neglected, we arrive at a bound for $P_m$,
	\begin{equation}
	P_m \leq \frac{2 \kappa R}{\eta-3\kappa+2\kappa \eta},
	\label{eq:pm}
	\end{equation}
	where the probability of non-empty pulses $R = p_1 + p_2 \geq p_1$ can be measured directly from the source. From our measurements, we find $\kappa = 1.1\times 10^{-5}$, $\eta = 0.06$, and $R = 0.033$, and with  \cref{eq:pm} we determine the probability of multi-photon emission to be $P_m \le 4.5 \times 10^{-6}$.

	\section{Weak coherent pulsed QKD}
	
	We compare the QD to a WCP source using the decoy-state BB84 protocol, which employs multiple intensity levels to counter photon number splitting attacks. The WCP source is realized using pulses from a Ti:sapph mode-locked laser, attenuated to a mean photon number  of $\mu = 0.5$ per signal pulse and $\nu = 0.1$ per decoy pulse. Note that because the optical pulses from a Ti:sapph laser have a phase relation, they are not directly suitable for secure QKD without phase randomization. (We omit this step for simplicity.)
	
	The estimated key length of decoy-state BB84 QKD with two intensity levels is
	\begin{equation}
	\label{eq:decoy-finite}
	L_\text{WCP} \geq nK_\mu q \left[ Y^L_1(1-H(E^U_1))-Q_\mu fH(E_\mu)-Q_\mu\Delta/n_\mu \right],
	\end{equation}
	where $n$ is the total number of transmitted pulses, $n_\mu$ is the number of detected signal pulses, $Q_\mu$ is the gain of the signal state, $Y_1^L$ is the lower bound of the single-photon gain, $E_1^U$ is the upper bound of the QBER with a correction for the finite-size effects on decoy state characterization \cite{curty2010}, $K_\mu = 0.9$ is a fraction of the pulses that are in a signal state, and $\Delta$ is a correction term for finite-size effects \cite{renner2005a,cai2009,scarani2008a}. Other parameters are as for the QD source.

	\section{Results and discussion}

	\begin{figure}
		\includegraphics{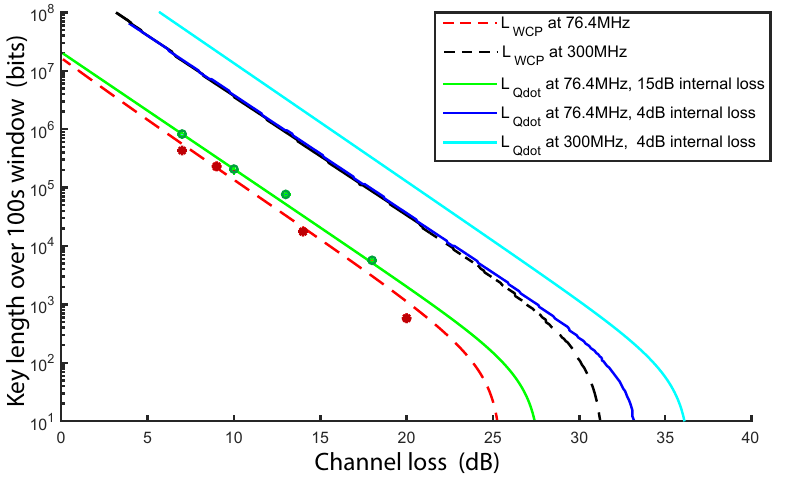}
		\caption{Secret key size over 100~s exchange, including finite-size effects, for varying channel loss. Symbols indicate experimental results, lines theoretical. Red circles and dashed line, WCP QKD with 76.4~MHz; green pluses and line, QD QKD with 76.4~MHz excitation frequency and $15~\text{dB}$ internal loss; blue line, extrapolation of 76.4~MHz QD QKD to $4.0~\text{dB}$ internal loss; black dashed line, extrapolation of WCP QKD to 300~MHz; light-blue line, extrapolation of QD QKD to 300~MHz excitation and $4.0~\text{dB}$ internal loss. }
		\label{fig:results}
	\end{figure}
	
	Experimental and theoretical results are shown in \cref{fig:results}. With an observed QBER of ${\approx} 2\%$, the QD QKD system can effectively tolerate channel loss up to ${\approx}27~\text{dB}$ (green line in \cref{fig:results}). 
	This tolerable loss of the QD QKD system is notably higher than the WCP protocol with the same repetition rate (red dashed line) despite its high internal loss of  $15~\text{dB}$.
	
	The performance of a QD QKD system will be further enhanced by realistic improvements to the source's internal losses. Other demonstrations of QD sources~\cite{gazzano2013,Bulgarini:2014aa} have reported an optimistic $4.0~\text{dB}$ (or $60\%$) internal loss, owing to a $50\%$ fiber coupling efficiency and an $80\%$ practical photon generation efficiency. Such improved internal coupling of the source brings the loss tolerance of the system up to $32~\text{dB}$ (blue line in \cref{fig:results}), even surpassing a decoy-state QKD system with 300~MHz repetition rate (black dashed line). Finally, assuming QD uses high-efficiency coupling and is also operated at the  300~MHz rate  (light blue line), our extrapolation predicts the QD QKD could tolerate significantly higher channel loss---close to $37$~dB---than a WCP QKD ($32$~dB), or generate up to one order of magnitude higher key length per 100~s satellite pass, significantly outperforming the WCP QKD system under the same channel conditions. The generation of the four QKD states from a QD emitter may introduce internal losses of 3 to 4~dB due to insertion losses from electro-optic modulators or passive coupling, and the key generation rates in would be proportionally lower. However, the advantage of QD QKD over WCP QKD still holds.

	\section{Conclusion}
	
	We  experimentally and theoretically compared the performance of QD and WCP QKD under finite-size effects for the purposes of secure communication in constrained channels such as with satellites. In particular, we devised a novel method to characterize and determine the upper bound of multi-photon emissions from a QD emitter, and include finite-size effects due to the limited link duration of about 100~s expected for a satellite contact.  Remarkably,  our results show that a QD QKD system operated at 76.4~MHz repetition rate, and despite $15~\text{dB}$ internal loss, outperforms a decoy-state WCP QKD running at the same repetition rate, especially at high channel transmission loss. The performance of QD QKD could be improved further by reducing the internal loss, and at an optimistic, but still practical $4.0~\text{dB}$ the performance advantage for a QD system is almost an order of magnitude better than WCP.  This QD sample was  driven by an off-resonant excitation scheme, and future work using resonant excitation could reduce the multi-photon emission even further as well as improve timing jitter and repetition rate of the emitted photon pulses. 
	
	State-of-the-art quantum dots coupled to microcavities have shown lifetimes of ${\approx}60~\pico\second$ \cite{Wang2019}, and can be driven on resonance by a GHz-rate pulsed laser. This is equivalent to the current clock rate used in high-speed WCP QKD \cite{gordon2005,dixon2008,wang2012}. Utilizing quantum dot sources in  satellite-uplink-based QKD is  very appealing, because the bulky components, such as the cryogenic system for the single-photon source, are located on the ground station. Note, however, the wavelength of single-photon pulses in this study (${\approx}890~\nano\meter$) is not optimal for the satellite uplink \cite{bourgoin2013}. Further study on other quantum dot materials that emit at better wavelengths, or the possibility of using frequency conversion of the quantum dot single-photon source, is needed.
	
	The narrow-band emission of QD is suited well for filters used in very light-polluted, or even daylight, environments. In addition, QD devices have the potential to generate entangled photon pairs, or even produce multiplexed emission, all of which could be helpful for interconnecting multiple users and enhancing the QKD rates in the future. Our theoretical analysis should be applicable to  QKD systems with other single-photon emitters, and we believe our study can help spark  interest in the advancements of QKD with true single-photon sources. As secure communication over long and global distances becomes more important than ever, the  enhancement of satellite QKD using true single-photon emitters is anticipated to have a major contribution to help make this happen. Our findings demonstrate that a quantum dot can indeed be a viable and beneficial photon emitter in a  QKD system. 
	
	\section{Acknowledgment}
	
	We thank N.~L{\"u}tkenhaus, J.P. Bourgoin, and J. Lin for discussions and technical support. This work was supported by the  Industry Canada, Canada Fund for Innovation, Ontario MRI, Ontario Research Fund, NSERC (programs Discovery, CryptoWorks21, Strategic Partnership Grant), and the Canadian Space Agency.  P.C.\ acknowledges support by Thai DPST scholarship.
	
	\def\bibsection{\medskip\begin{center}\rule{0.5\columnwidth}{.8pt}\end{center}\medskip} 

	\bibliography{library}

\end{document}